\documentclass[conference]{IEEEtran}

\usepackage{cite}
\usepackage{amsmath,amssymb,amsfonts}
\usepackage{algorithmic}
\usepackage{graphicx}
\usepackage{textcomp}
\usepackage{float}  
\usepackage{xcolor}
\usepackage{makecell}
\def\BibTeX{{\rm B\kern-.05em{\sc i\kern-.025em b}\kern-.08em
    T\kern-.1667em\lower.7ex\hbox{E}\kern-.125emX}}
\IEEEoverridecommandlockouts

\makeatletter
\def\ps@IEEEtitlepagestyle{
  	\def\@oddfoot{\Footer} % THIS PLACES THE FOOTER ON THE FIRST PAGE
	}
\let\old@ps@headings\ps@headings
\let\old@ps@IEEEtitlepagestyle\ps@IEEEtitlepagestyle
\def\confheader#1{%
  \def\ps@headings{%
    \old@ps@headings%
    \def\@oddhead{\strut\hfill#1\hfill\strut}%
    \def\@evenhead{\strut\hfill#1\hfill\strut}%
    \def\@oddfoot{\Footer} % THIS PLACES THE FOOTER ON ALL PAGES AFTER THE FIRST PAGE
  }%
  \def\ps@IEEEtitlepagestyle{%
    \old@ps@IEEEtitlepagestyle%
    \def\@oddhead{\strut\hfill#1\hfill\strut}%
    \def\@evenhead{\strut\hfill#1\hfill\strut}%
  }%
  \ps@headings%
}
\makeatother
\confheader{} % HEADER EXAMPLE 
\def\Footer{
  {\footnotesize
 \begin{minipage}{\textwidth}
 \centering \scriptsize{ %Approval ID: $<$XXXX-2025-XXXX$>$. (MARKED FOR DEMO PURPOSES ONLY)\\
 }  \end{minipage}
 }
}
    
\begin{document}

\title{Causal AI For AMS Circuit Design: Interpretable Parameter Effects Analysis
}

\author{\IEEEauthorblockN{Mohyeu Hussain}
\IEEEauthorblockA{\textit{University of Florida} \\
Email: mohyeuhussain@ufl.edu}
\and
\IEEEauthorblockN{David Koblah}
\IEEEauthorblockA{\textit{University of Florida} \\
Email: dkoblah@ufl.edu}
\and
\IEEEauthorblockN{Reiner Dizon-Paradis}
\IEEEauthorblockA{\textit{University of Florida} \\
Email: reinerdizon@ufl.edu}
\and
\IEEEauthorblockN{Domenic Forte}
\IEEEauthorblockA{\textit{University of Florida} \\
Email: dforte@ece.ufl.edu}

}

\maketitle

\begin{abstract}
Analog‑mixed‑signal (AMS) circuits are highly nonlinear and operate on continuous real‑world signals, making them far more difficult to model with data‑driven AI than digital blocks. To close the gap between structured design data (device dimensions, bias voltages, etc.) and real‑world performance, we propose a causal‑inference framework that first discovers a directed‑acyclic graph (DAG) from SPICE simulation data and then quantifies parameter impact through Average Treatment Effect (ATE) estimation. The approach yields human‑interpretable rankings of design knobs and explicit ``what‑if'' predictions, enabling designers to understand trade‑offs in sizing and topology. We evaluate the pipeline on three operational‑amplifier families (OTA, telescopic, and folded‑cascode) implemented in TSMC 65nm and benchmark it against a baseline neural‑network (NN) regressor. Across all circuits the causal model reproduces simulation‑based ATEs with an average absolute error of less than 25\%, whereas the neural network deviates by more than 80\% and frequently predicts the wrong sign. These results demonstrate that causal AI provides both higher accuracy and explainability, paving the way for more efficient, trustworthy AMS design automation.
\end{abstract}

\section{Introduction} \label{sec:intro}
Analog circuits are the backbone of virtually every electronic system, from smartphones and medical‑imaging equipment to radar, satellite communications, and weapons‑control subsystems used in government and defense applications. 
As integrated circuits (ICs) become ever more pervasive, the demand for high‑performance analog components has risen sharply, spurring intense interest in automating analog design.

Despite this ubiquity, designers still confront three dominant pain points that limit productivity and yield.
\begin{enumerate}

\item \textbf{Manual design bottlenecks and expert dependency.} Engineers must rely on hand‑tuned transistor sizing and iterative SPICE sweeps; a typical full‑chip analog block can require more than ten SPICE runs per iteration, consuming several days of compute time and often delaying project schedules by 2–3 weeks~\cite{synopsysSynopsysLeads}.\textbf{}

\item \textbf{Fragile designs across process‑voltage‑temperature (PVT) corners and post‑layout variability.} Extensive Monte‑Carlo analysis—often hundreds of corners—is needed to guarantee performance, stretching verification cycles to months if not accelerated~\cite{synopsysSynopsysLeads}. This variability is a major source of re‑spins and design failures.

\item \textbf{Opaque trade‑offs (e.g., gain vs. bandwidth, noise vs. power, linearity vs. swing).} Because these compromises remain hidden in black‑box simulations, designers must manually explore multidimensional design spaces, a task that typically consumes 30–40\% of the overall design effort~\cite{miscircuitosGainBandwidthTradeOff}.
\end{enumerate}
These quantitative constraints highlight why AI models must be both accurate and explainable, so that design decisions can be traced, justified, and refined by human experts.

Unlike digital blocks, analog circuits exhibit continuous, nonlinear behavior, making accurate modeling and synthesis especially challenging. 
Their performance depends on device‑level characteristics that vary with PVT, requiring designers to capture subtle relationships across a wide operating range. 
Moreover, the lack of standardized design primitives and the strong coupling between circuit blocks further complicate the creation of reusable, automated flows~\cite{mina2022review}.

Bridging the gap between the abstract data structures that capture device geometry and functionality and the circuit’s actual behavior is a prerequisite for effective AI‑driven design. 
Equally important is the development of AI models that are not only precise but also explainable and interpretable, ensuring that design decisions remain consistent with expert knowledge and can be readily understood by users.
Recent work in analog‑circuit automation often treats a circuit as a black box, ignoring the underlying reasoning~\cite{shi2018toward}. 
Yet understanding how a circuit operates -- identifying the most influential parameters, uncovering causal pathways that shape performance, and tracing the step‑by‑step design rationale -- is essential for robust analog design~\cite{jiao2015analog,jiao2017modeling}.
By constructing models that emulate the design workflow of analog‑mixed‑signal (AMS) engineers, we can harness AI to both deepen insight into device behavior and enhance the optimization of AMS components.

In this study, we adopt a causal inference framework for analog circuit design that mitigates the traditional trial‑and‑error workflow and delivers human‑interpretable guidance for parameter optimization. 
Causal AI explicitly models cause‑and‑effect relationships, allowing one to infer how changes in a design variable propagate to performance metrics rather than merely learning statistical correlations. 
Because statistical correlations can arise from hidden confounding variables or coincidental patterns, they may suggest a strong link between a parameter and a metric that vanishes when the underlying cause is controlled, leading designers to pursue ineffective or counter‑productive adjustments. 
By extracting causal graphs directly from simulation data, the framework identifies the parameters that most strongly drive outcomes such as gain, phase margin, or power consumption, while de-emphasizing those with negligible influence. 
Treatment effects are quantified using Average Treatment Effect (ATE) estimation, and the resulting causal model is benchmarked against conventional neural network predictors and raw simulation results. 
Consequently, the approach delivers both high predictive accuracy and transparent interpretability, showing not only \textit{which} parameters matter but also \textit{how} they affect the design objectives.

To the best of our knowledge, this work constitutes the first application of causal inference to analog circuit design. Our main objective is to establish a causal inference-driven paradigm for analog circuit design where, given circuit simulation data, the framework delivers:
    \begin{itemize}
        \item A ranked set of design parameters that most strongly influence the targeted performance metrics.
        \item Human-interpretable insights on tradeoffs, bottlenecks, and opportunities for refinement, closely aligned with simulation results but more efficient than trial-and-error design.
    \end{itemize}
The rest of the paper is organized in the following manner.
In Section~\ref{sec:background}, we describe explainable causal modeling for analog circuits, related works and some necessary terms. 
Section~\ref{sec:methods} introduces our proposed framework. 
In Section~\ref{sec:results}, we discuss our experiment results and our analysis on them and finally Section~\ref{sec:conclusion} concludes the paper. 

\section{Background} \label{sec:background}

In this section, we first discuss why interpretability is crucial for analog‑circuit design and how causal discovery can provide a transparent model of the design space. Then, we introduce the Average Treatment Effect (ATE) as the quantitative measure used to assess the impact of individual design parameters in this paper. 

\subsection{Explainable Causal Modeling for Analog Circuits}
\label{subsec:xai}

Interpretability is essential in analog design because engineers must justify sizing choices, corner‑case analyses, and trade‑off decisions to both internal reviewers and external regulators.  
An explainable method therefore has to reveal how circuit parameters influence key performance indicators (KPIs), not merely predict the KPI values.

Early attempts at explainability focused on symbolic reasoning.  
Shi \emph{et al.}~\cite{shi2018toward} introduced computer‑aided design reasoning (CAR), but the approach is limited to handcrafted symbolic manipulation and does not leverage modern AI/ML techniques. 
More recent work by Abuelnasr \emph{et al.}~\cite{abuelnasr2021causal} combined neural networks with various ML models to predict causal relations, yet pure prediction models cannot capture true cause‑effect dependencies needed for trustworthy interpretation.

A principled way to obtain those dependencies from data is causal discovery. 
Causal discovery algorithms infer a directed acyclic graph (DAG) that encodes the causal influence among design variables directly from simulation data. 
The main families of methods are:
\begin{itemize}
  \item \textbf{Constraint‑based} approaches (e.g., PC, FCI) that test conditional independencies to prune edges;
  \item \textbf{Score‑based} approaches (e.g., GES) that search for the graph maximizing a penalized likelihood score;
  \item \textbf{Hybrid} methods that combine the two paradigms to improve scalability on high‑dimensional circuit data.
\end{itemize}

In the analog domain, Jiao \emph{et al.}~\cite{jiao2015analog,jiao2017modeling,jiao2016three} demonstrated that signal‑flow‑graph‑driven causal models can automatically identify the parameters that dominate performance in op‑amp families. 
Their experiments showed that causal modeling isolates the most influential variables for gain, bandwidth, slew‑rate and noise, and can guide the sizing loop more efficiently than blind trial‑and‑error.  
However, those studies were limited to a handful of circuits and a single technology node, which hampers generalizability.

Our framework builds on these insights by applying a modern hybrid causal‑discovery pipeline to large simulated design spaces across multiple circuit topologies and technology corners.  
The resulting directed acyclic graph (DAG) provides a transparent, engineer‑readable explanation of the design space: edges point from a design variable (the \textit{treatment}) to the performance metric (the \textit{outcome}), while the graph also makes explicit the confounding variables and colliders that can otherwise induce spurious correlations. 
Consequently, the causal graph serves as the backbone for the quantitative impact analysis described in the next subsection.

\subsection{Average Treatment Effect (ATE)}
\label{subsec:ATE}
The Average Treatment Effect or ATE quantifies the expected change in a performance metric when a design parameter is intervened upon, while all other variables retain their observed distribution.

\vspace{0.5ex}

\noindent \textbf{Expectation \(\mathbb{E}[\cdot]\).}
For a random variable \(Z\), the expectation (i.e., the mean) with respect to its probability distribution is
\[
\mathbb{E}[Z] \;=\;
\begin{cases}
\displaystyle\sum_{z} z\,p_{Z}(z), & \text{discrete }Z,\\[6pt]
\displaystyle\int z\,p_{Z}(z)\,dz, & \text{continuous }Z,
\end{cases}
\]
where \(p_{Z}(z)\) denotes the probability mass (or density) function of \(Z\).

\vspace{0.5ex}

\noindent \textbf{Intervention \(\operatorname{do}(\cdot)\).}
The \(\operatorname{do}(\cdot)\) operator, introduced by Pearl~\cite{neuberg2003causality}, represents an external manipulation that forces a variable to a specified value, breaking any incoming causal links.  In other words,
\[
\operatorname{do}(T=t)
\]
means “set the treatment variable \(T\) to the value \(t\) by intervention, irrespective of the natural causes that would otherwise determine \(T\).”  This distinguishes true causal effects from mere observational correlations.

\vspace{0.5ex}

\noindent \textbf{ATE Formalism.}
With a binary treatment \(T\) (e.g., increasing a bias current) and an outcome \(Y\) (e.g., AC gain), the ATE is
\[
\text{ATE}
   = \mathbb{E}\!\bigl[\,Y \mid \operatorname{do}(T=1)\,\bigr]
     - \mathbb{E}\!\bigl[\,Y \mid \operatorname{do}(T=0)\,\bigr].
\]

In our workflow, the ATE is estimated from the learned causal graph using propensity‑score‑adjusted regression (or, equivalently, by applying do‑calculus on the DAG).  
This yields an interpretable “impact score’’ for each design variable that can be directly compared against the predictions of a baseline neural‑network model, which would otherwise rely only on observed correlations.

\section{Proposed Causal Analysis Methodology} \label{sec:methods}

\begin{figure}[t]
\centering
\includegraphics[width=0.4\textwidth]{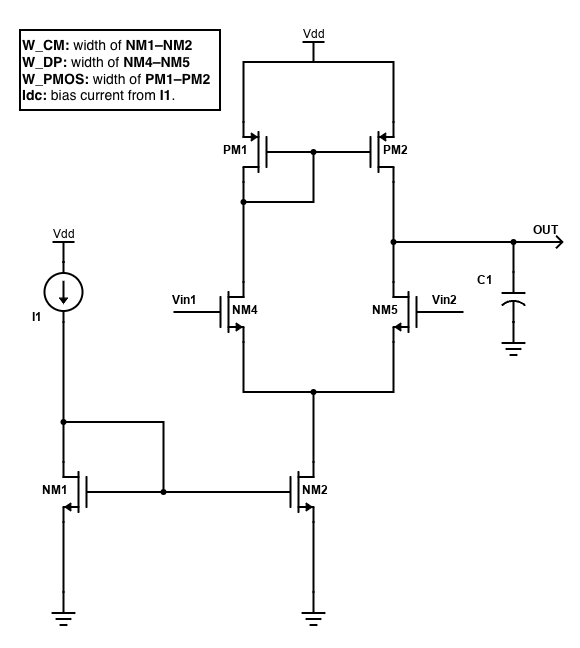}
% \vspace{0.1in}
\caption[]{\centering Schematic of OTA with current mirror.}
\label{fig:OTA with CM}
% \vspace{0.25in}
\end{figure}

Figure~\ref{fig:OTA with CM} shows a typical operational‑transconductance amplifier (OTA).  
Its performance is governed by a set of design variables -- transistor width‑to‑length (W/L) ratios, bias voltages, and bias currents -- that are traditionally tuned by repeatedly running SPICE simulations and manually adjusting the parameters.  
Although experienced designers can rely on intuition to decide which knobs to turn in order to meet specifications such as gain or phase margin, this trial‑and‑error process becomes increasingly labor‑intensive and error‑prone as the circuit topology grows in complexity.  

A causal model mitigates these difficulties by explicitly uncovering the cause‑effect links between design variables and performance metrics.  
The resulting causal graph highlights the parameters that exert the strongest influence, exposes performance bottlenecks, and reveals how distinct circuit blocks interact to satisfy design constraints.  
Consequently, designers gain a concise, interpretable roadmap for sizing, can explore new topologies or refine existing ones with far fewer simulation cycles, and achieve more efficient, data‑driven optimization of the OTA.

\begin{figure}[t]
\centering
\includegraphics[width=0.5\textwidth]{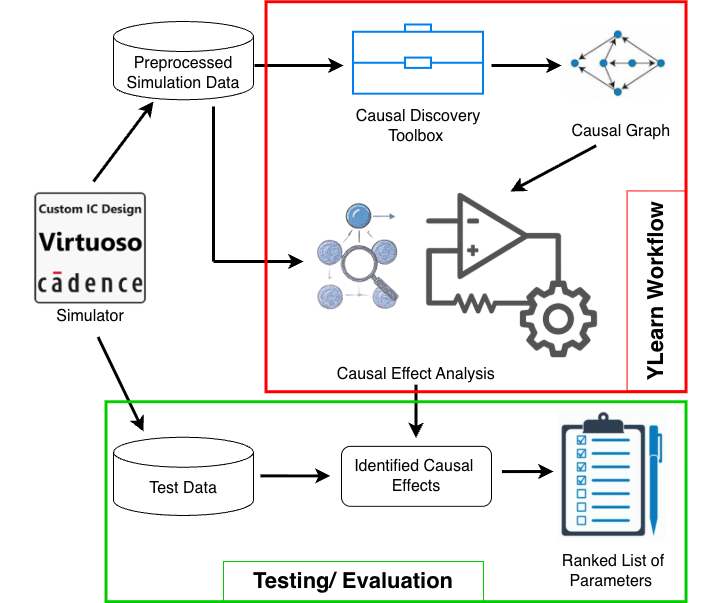}
% \vspace{0.01in}
\caption[]{Overview of the proposed methodology. Preprocessed simulation data are used to train a causal model and estimate causal effects using YLearn (red, training stage). The learned effects are then applied to held‑out test data and evaluated in a Python pipeline (green, testing stage) to produce a ranked list of influential parameters.}
\label{fig:methodology}
% \vspace{0.25in}
\end{figure}
%\vspace{-0.05in}

A schematic of the entire workflow is shown in Figure~\ref{fig:methodology}.
Our framework begins by generating simulation data for the circuits using Cadence Virtuoso Spectre simulator and removing any rows or columns with missing outcomes, as well as unnecessary columns generated during the simulation. Next, the preprocessed simulation data is fed into the causal-discovery toolbox. 
The toolbox extracts a DAG that encodes the causal relationships among all design variables and performance metrics.  
In this graph, each edge indicates a direct influence, while the downstream topology reveals indirect (mediated) effects on the outcomes.
The resulting causal graph serves as the backbone for quantitative effect analysis.

To convert the graph into actionable numbers, we employ YLearn~\cite{YLearn}, an open‑source Python library for causal inference.  
YLearn implements a hybrid discovery pipeline (constraint‑based skeleton followed by a score‑based refinement) and provides a suite of effect‑estimation tools, including ATE and Individual Treatment Effect (ITE).  
Using YLearn, we estimate how variations in key design parameters -- e.g., the widths of differential‑pair and PMOS transistors (\texttt{W\_DP}, \texttt{W\_PMOS}), the bias current (\texttt{Idc}), and channel length (\texttt{L}) -- affect performance targets such as AC gain, phase margin, or bandwidth.

Because the effect estimates are derived from the same causal graph uncovered in the discovery stage, they are interpretable (the adjustment set is explicit) and consistent with the underlying SPICE simulations.  
This contrasts with purely black‑box predictors, which can achieve high accuracy but offer no insight into why a particular parameter is important.  
Consequently, our YLearn‑based pipeline yields a reliable, data‑driven design assistant that can prioritize the most impactful knobs, expose hidden bottlenecks, and streamline the sizing process.

\section{Results and Discussions} \label{sec:results}
In this section we evaluate the proposed causal‑inference pipeline on three representative analog blocks: Operational Transconductance Amplifier (OTA), Telescopic Op Amp, and Folded Cascode Op Amp. 
 We compared our models' ATE against both a baseline neural network (NN) model and direct simulation results. The causal model is implemented using the Why framework and consists of Double Machine Learning, that includes a Random Forest x-model, an MLP-based y-model with a single hidden layer, and an ElasticNet yx-model. The baseline Neural Network (NN) model is implemented as an S-learner using a fully connected multilayer perceptron with two hidden layers of 128 and 64 neurons. 
 The simulation results were generated using Cadence Virtuoso on the TSMC 65nm technology node. For the OTA and the telescopic op amp, we used 20,000 simulation samples, whereas for the folded cascode op amp, we used 38,000 simulation samples. We used the same channel length \texttt{L} for all transistors across all circuits, and during the experiments the transistor lengths were varied simultaneously.
 
 \subsection{Operational Transconductance Amplifier (OTA)}

 To evaluate our model on real world circuits, we started with OTA with current mirror (Figure~\ref{fig:OTA with CM}). 

\begin{figure}[t]
\centering
\includegraphics[width=0.4\textwidth]{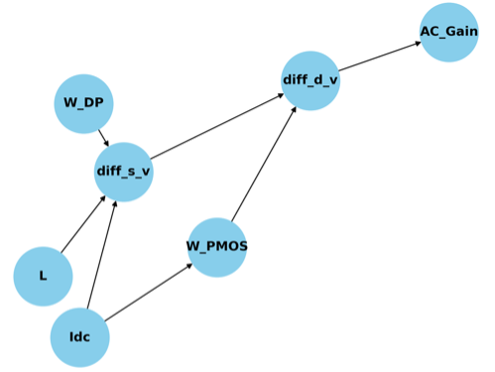}
% \vspace{0.1in}
\caption[]{\centering Causal discovery graph of OTA.}
\label{fig:discovery}
% \vspace{0.25in}
\end{figure}

\noindent \textbf{Causal Graph.} Figure~\ref{fig:discovery} displays the DAG produced by the YLearn discovery stage for the OTA.
The width of the differential pair (\texttt{W\_DP}) and the width of the load PMOS (\texttt{W\_PMOS}) feed directly into the intermediate node voltages \texttt{diff\_d\_v} and \texttt{diff\_s\_v}, which in turn drive the output \texttt{AC\_Gain}.
This structure makes explicit which design knobs have direct versus mediated influence on the performance metric.

 \vspace{0.5ex}

\begin{figure}[t]
\centering
\includegraphics[width=0.47\textwidth]{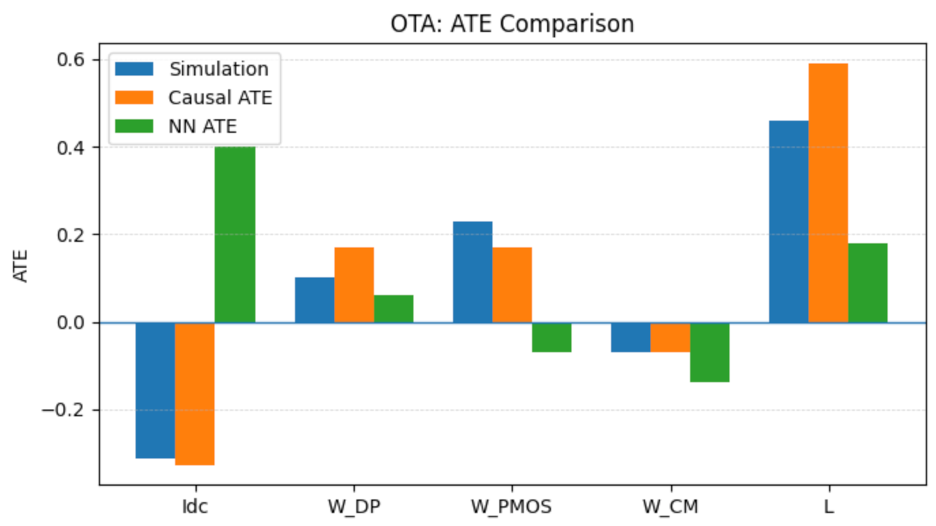}
% \vspace{0.1in}
\caption[]{\centering Comparison of ATE values for OTA.}
\label{fig:OTAgraph}
% \vspace{0.25in}
\end{figure}

\noindent \textbf{Quantitative ATE comparison.} Table~\ref{tab:ota} and Figure~\ref{fig:OTAgraph} summarize the ATE values for a 10\% perturbation of each parameter, together with the percent deviation from the simulation baseline.
The causal model reproduces the SPICE ATEs with an average absolute difference of 25.8\%, whereas the NN regressor deviates by 111.8\%.  
In several cases (e.g., \texttt{Idc}, \texttt{W\_CM}), the NN even predicts the wrong sign, indicating a confounding‑driven bias.

Because the causal graph isolates the true pathways from \texttt{W\_DP} and \texttt{W\_PMOS} to \texttt{AC\_Gain}, a designer can focus sizing effort on these two transistors and on the bias current \texttt{Idc}.  The NN’s misranking (e.g., placing \texttt{W\_CM} above \texttt{W\_PMOS}) would lead to unnecessary simulation cycles.

\begin{table}[t]
\caption{Comparison of ATE results for OTA}
\label{tab:ota}
\begin{center}
\begin{tabular}{|c|c|c|c|}
\hline
\textbf{\emph{Parameter}} &
\makecell{\textbf{\emph{Simulation Data}} \\ \textbf{\emph{ATE}}} &
\makecell{\textbf{\emph{(\%) Diff. with}} \\ \textbf{\emph{Causal ATE}}} &
\makecell{\textbf{\emph{(\%) Diff. with}} \\ \textbf{\emph{NN ATE}}} \\
\hline
\texttt{Idc} & -0.313 & 4.47\% & -227.8\%\\
\hline
\texttt{W\_DP} & 0.1 & 70\% & -40\%\\
\hline
\texttt{W\_PMOS} & 0.23 & -26\% & -130.4\%\\
\hline
\texttt{W\_CM} & -0.07 & 0\% & 100\%\\
\hline
\texttt{L} & 0.46 & 28.26\% & -60.87\%\\
\hline
\multicolumn{2}{|c|}{\textbf{Avg. of Absolute Values}} & \textbf{25.75\%} & \textbf{111.81\%} \\
\hline
\end{tabular}
\end{center}
\end{table}

\subsection{Telescopic Op Amp}

\begin{figure}[t]
\centering
\includegraphics[width=0.35\textwidth]{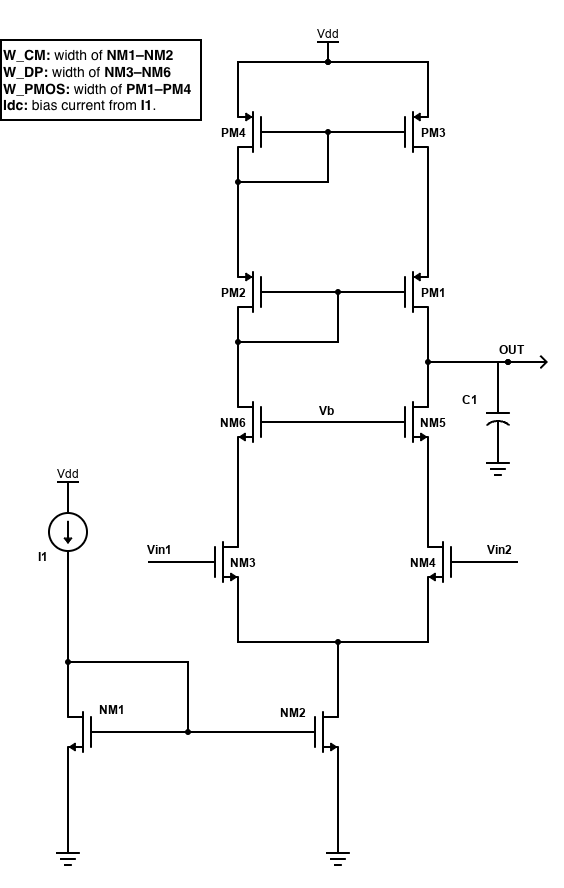}
% \vspace{0.1in}
\caption[]{\centering Schematic of a single-ended telescopic op amp.}
\label{fig:telescopic}
% \vspace{0.25in}
\end{figure}

Figure~\ref{fig:telescopic} shows the circuit for the telescopic op amp. 

\vspace{0.5ex}

\noindent \textbf{Causal Graph.} The corresponding DAG (omitted for space) shows a direct edge from \texttt{W\_PMOS}, \texttt{Idc} and \texttt{L} to the output node whereas the other parameters showed mediated influence. 

 \vspace{0.5ex}

\noindent \textbf{Quantitative ATE comparison.} Table~\ref{tab:telescopic} and Figure~\ref{fig:telescopic} report results from a similar 10\% perturbation study as the OTA. 
The causal model's average absolute deviation is 24.1\%, compared with 85.9\% for the NN. 
For the most influential knob \texttt{Idc} the NN even flips the sign (+1.13 versus –0.93 in simulation), while the causal estimate (–0.924) stays within 1\% of the ground truth.

The causal graph correctly predicts that \texttt{W\_PMOS} dominates the gain while \texttt{Idc} governs the low‑frequency pole.  
The NN’s weaker ranking suggests that it is unable to disentangle the confounding between the tail‑current bias and the differential‑pair sizing, a problem that worsens as the circuit topology becomes more intricate.

\begin{table}[t]
\caption{Comparison of results for Telescopic Op Amp}
\label{tab:telescopic}
\begin{center}
\begin{tabular}{|c|c|c|c|}
\hline
\textbf{\emph{Parameter}} &
\makecell{\textbf{\emph{Simulation Data}} \\ \textbf{\emph{ATE}}} &
\makecell{\textbf{\emph{(\%) Diff. with}} \\ \textbf{\emph{Causal ATE}}} &
\makecell{\textbf{\emph{(\%) Diff. with}} \\ \textbf{\emph{NN ATE}}} \\
\hline
\texttt{Idc} & -0.93 & 0\% & -221.5\%\\
\hline
\texttt{W\_DP} & -0.65 & 9.23\% & 29.23\%\\
\hline
\texttt{W\_PMOS} & 1.84 & -35.33\% & -65.22\%\\
\hline
\texttt{W\_CM} & -0.25 & 16\% & 20\%\\
\hline
\texttt{L} & -0.8 & 60\% & 93.75\%\\
\hline
\multicolumn{2}{|c|}{\textbf{Avg. of Absolute Values}} & \textbf{24.11\%} & \textbf{85.94\%} \\
\hline
\end{tabular}
\end{center}
\end{table}

\begin{figure}[t]
\includegraphics[width=0.47\textwidth]{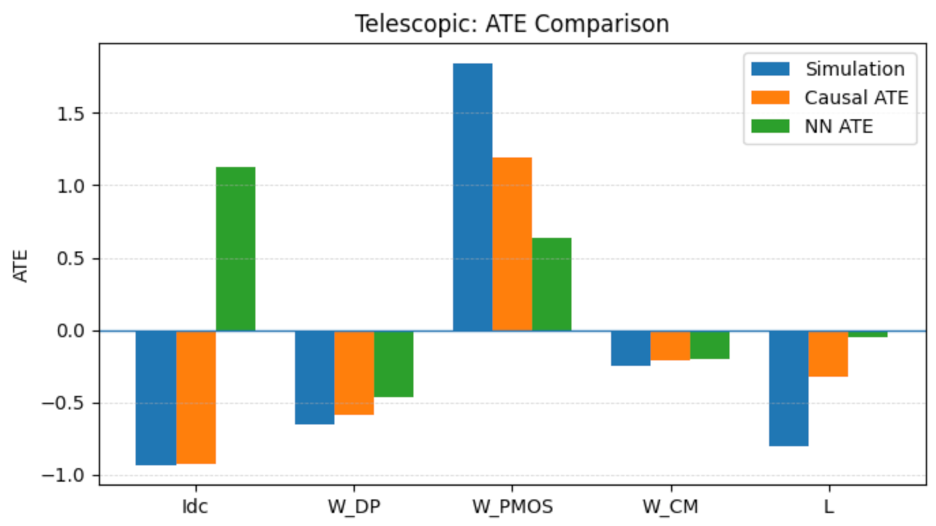}
% \vspace{0.1in}
\caption[]{\centering Comparison of ATE values for telescopic op amp.}
\label{fig:telescopicgraph}
% \vspace{0.25in}
\end{figure}

\subsection{Folded Cascode Op Amp}

\begin{figure}[t]
\centering
\includegraphics[width=0.5\textwidth]{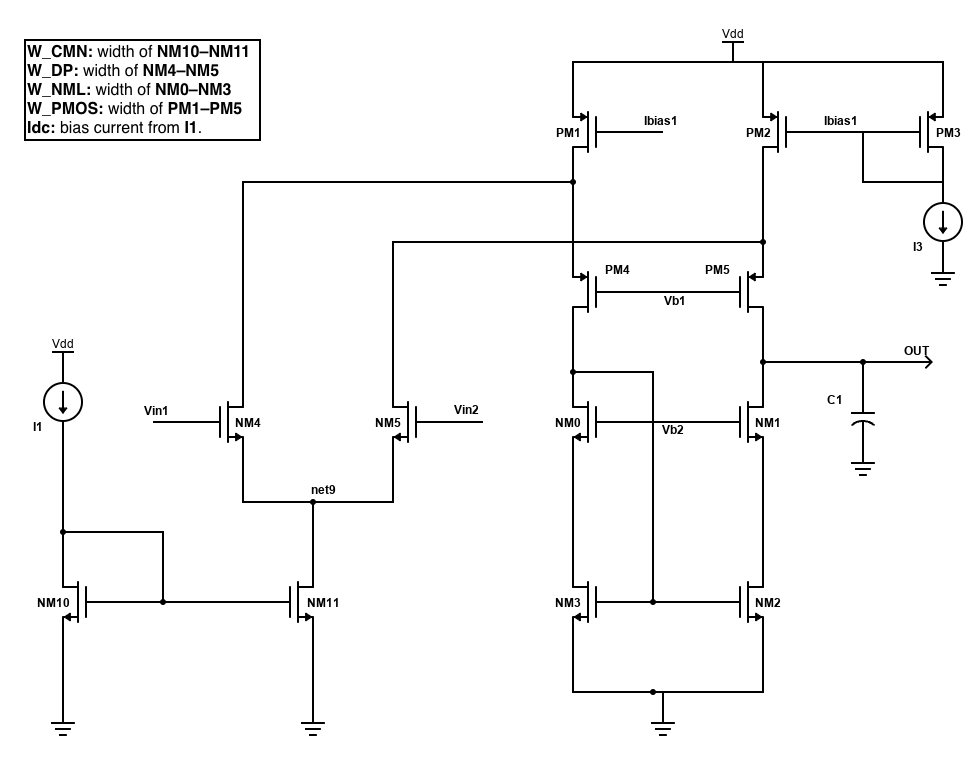}
% \vspace{0.1in}
\caption[]{\centering Schematic of a single-ended folded cascode op amp.}
\label{fig:folded}
% \vspace{0.25in}
\end{figure}

Lastly, to prove the superiority of our model than the NN model, we experimented on a folded cascode op amp which has 13 transistors (see Figure~\ref{fig:folded}), being the most complex one among all the op amps discussed so far. We examined 5 different parameters. 

\vspace{0.5ex}

\noindent \textbf{Causal Graph.} The DAG (not shown for space) reveals the input pair (\texttt{W\_DP}), the load branch (\texttt{W\_PMOS}), (\texttt{W\_NML}) and the current‑mirror bias (\texttt{Idc}), all contribute to the gain directly. As circuit complexity increases, more parameters shows mediated influence, and this circuit also shows several such parameters affecting the gain. 

\vspace{0.5ex}

\noindent \textbf{Quantitative ATE comparison.} Table~\ref{tab:folded} and Figure~\ref{fig:folded} summarize the 10\% perturbation results.  The causal model achieves an average absolute deviation of only 7.6\%, whereas the NN’s average deviation blows up to 237.7\%. The NN even predicts a positive ATE for \texttt{Idc} (+3.2) while the true effect is strongly negative (–0.45).

\begin{table}[t]
\caption{Comparison of results for Folded Cascode Op Amp}
\label{tab:folded}
\begin{center}
\begin{tabular}{|c|c|c|c|}
\hline
\textbf{\emph{Parameter}} &
\makecell{\textbf{\emph{Simulation Data}} \\ \textbf{\emph{ATE}}} &
\makecell{\textbf{\emph{(\%) Diff. with}} \\ \textbf{\emph{Causal ATE}}} &
\makecell{\textbf{\emph{(\%) Diff. with}} \\ \textbf{\emph{NN ATE}}} \\
\hline
\texttt{Idc} & -0.45 & -6.67\% & 811.11\%\\
\hline
\texttt{W\_DP} & 0.27 & 18.52\% & -170.37\%\\
\hline
\texttt{W\_PMOS} & 0.19 & -5.26\% & -63.16\%\\
\hline
\texttt{W\_NML} & -0.18 & -3.89\% & 33.33\%\\
\hline
\texttt{L} & -0.28 & -3.57\% & 110.71\%\\
\hline
\multicolumn{2}{|c|}{\textbf{Avg. of Absolute Values}} & \textbf{7.58\%} & \textbf{237.73\%} \\
\hline
\end{tabular}
\end{center}
\end{table}

\begin{figure}[t]
\includegraphics[width=0.47\textwidth]{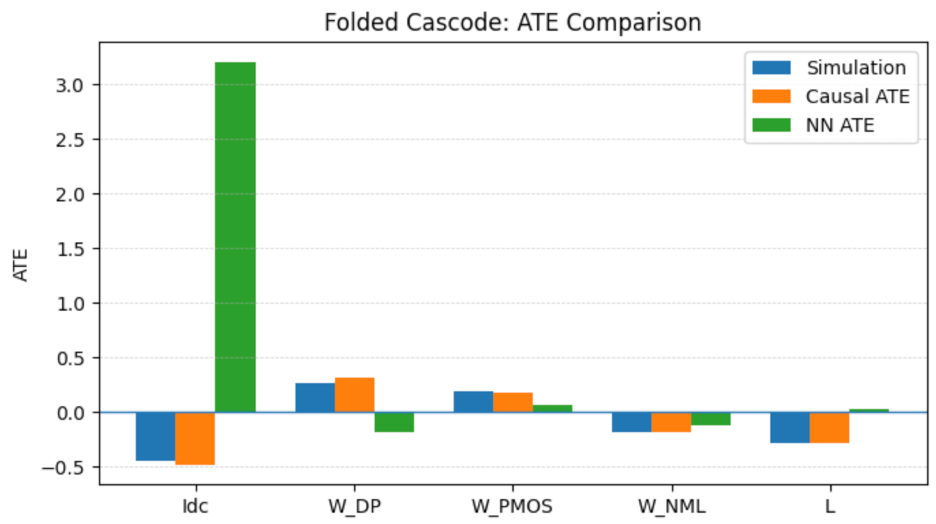}
% \vspace{0.1in}
\caption[]{\centering Comparison of ATE values for folded cascode op amp.}
\label{fig:foldedgraph}
% \vspace{0.25in}
\end{figure}

\subsection{Discussion}
If we look closely at and compare the results for all the op amps, we can find some insightful behavior. 

\begin{itemize}

\item \textbf{Scalability of causal inference.}  As circuit size grows (13 transistors in folded cascode vs. 7 in the OTA),
the causal model’s error actually decreases (7.6\% vs. 25.8\%). 
This reflects the graph‑based approach’s ability to ``explain away'' spurious correlations that would otherwise pollute a purely data‑driven NN.

\item \textbf{Bias from hidden confounders.}  The NN systematically misinterprets the effect of \texttt{Idc}, a variable that appears in many paths.  
Because the NN treats all features as independent predictors, it cannot separate the direct current‑bias effect from the indirect influence via node voltages -- exactly the decomposition that the causal DAG provides.

\item \textbf{Ranking of knobs.} Across all three circuits, the causal analysis consistently places \texttt{Idc} (bias current) and the load‑device width (\texttt{W\_PMOS} or \texttt{W\_NML}) at the top of the impact list, matching well‑known design intuition.
The NN often swaps these rankings, which could mislead an engineer into an inefficient optimization loop.
\end{itemize}

The experiments demonstrate that a hybrid causal‑discovery plus ATE estimation pipeline yields accurate, interpretable effect sizes even for relatively complex analog blocks. 
Compared with a baseline NN regressor, the causal approach reduces the average absolute deviation from the simulation ground truth by a factor of $\approx$5 (25.8\% to 7.6\%) and eliminates sign‑flips that would otherwise cause design instability.

These findings have two practical ramifications. 
First, in design‑space pruning, engineers can immediately drop parameters whose
estimated ATE is negligible, cutting down the number of SPICE sweeps required for sizing.
Second, because the causal model explicitly conditions on the adjustment set identified in the DAG, it can answer ``what‑if'' questions (e.g., \textit{what gain would I obtain if I double \texttt{W\_PMOS} while keeping all other
variables fixed?}).
The adjustment set is the smallest collection of variables that blocks every backdoor path between the treatment (\texttt{W\_PMOS}) and the outcome (\texttt{AC\_Gain}), thereby removing spurious correlations caused by hidden confounders such as the bias current \texttt{Idc} or the transistor lengths. 
By fixing these covariates at their observed values, the model isolates the direct causal effect of the chosen parameter, whereas the plain NN regressor inadvertently attributes the influence of the confounders to \texttt{W\_PMOS}, often producing biased or even sign‑flipped predictions.

\section{Conclusion and Future Work} \label{sec:conclusion}

Analog‑circuit design is becoming increasingly demanding as technology nodes shrink and circuit topologies grow in complexity. 
In this context, explainability -- the ability to trace a designer’s reasoning and understand how design decisions affect specifications -- is essential for building trustworthy, data‑driven design tools. 
We have presented a causal‑inference framework that (i) discovers a DAG describing the cause‑and‑effect relationships among transistor dimensions, bias voltages, and performance metrics; (ii) uses the DAG to compute ATEs that rank design parameters by their true impact on specifications; and (iii) delivers these rankings in an interpretable form that highlights bottlenecks and opportunities for refinement. 
Experiments on three representative op‑amp families show that the causal model reproduces the ground‑truth ATEs from SPICE simulations with an average absolute error of less than 25\%, whereas a baseline neural‑network regressor deviates by more than 80\% on the same tasks. 
Beyond the quantitative improvements, the proposed methodology offers a practical design workflow: engineers can generate a compact causal graph from a modest set of simulations, identify the most influential knobs, and focus subsequent sizing or optimization cycles on those knobs, thereby reducing the number of costly SPICE sweeps required for convergence. 

Future work will extend the pipeline to circuits that incorporate common‑mode feedback (CMFB) and to larger mixed‑signal blocks. We plan to investigate hierarchical partitioning of the circuit netlist, building separate causal sub‑graphs for each partition and then stitching them together, which should preserve interpretability while scaling to full ASIC‑level designs.

\section*{Acknowledgment} \label{sec:acknowledgments}

This research was supported by the Semiconductor Research Corporation (SRC) under Task \#3160.037.

%\balance
\bibliographystyle{IEEEtran}
\bibliography{references}

\end{document}